\documentclass[runningheads]{llncs}

\usepackage{graphicx}
\usepackage{booktabs}
\usepackage{multirow}
\usepackage{makecell}
\usepackage{amssymb}
\usepackage{mathtools}
\usepackage[bb=pazo]{mathalpha}
\usepackage{array}
\usepackage{enumitem}
\usepackage[table]{xcolor}
\usepackage{soul}
\usepackage{arydshln}
\usepackage{hyperref}

\newcolumntype{H}{>{\setbox0=\hbox\bgroup}c<{\egroup}@{}}

\definecolor{lightgray}{RGB}{225, 225, 225}

\newcolumntype{M}[1]{>{\centering\arraybackslash}m{#1}}

\begin{document}

\title{Robustness in Fairness against Edge-level Perturbations in GNN-based Recommendation}
\titlerunning{Robustness in Fairness in Recommendation}

\author{%
Ludovico Boratto\inst{1}\orcidID{0000-0002-6053-3015} \and
Francesco Fabbri\inst{2}\orcidID{0000-0002-9631-1799} \and
Gianni Fenu\inst{1}\orcidID{0000-0003-4668-2476} \and
Mirko Marras\thanks{Corresponding author.}$^{,}$\inst{1}\orcidID{0000-0003-1989-6057} \and
Giacomo Medda\inst{1}\orcidID{0000-0002-1300-1876}
}

\authorrunning{Boratto et al.}

\institute{University of Cagliari, Cagliari, Italy\\
\email{\{ludovico.boratto,mirko.marras\}@acm.org, \{fenu,giacomo.medda\}@unica.it}, \and
Spotify, Barcelona, Spain\\
\email{francescof@spotify.com}}

\maketitle            

\begin{abstract}
Efforts in the recommendation community are shifting from the sole emphasis on utility to considering beyond-utility factors, such as fairness and robustness.
Robustness of recommendation models is typically linked to their ability to maintain the original utility when subjected to attacks.
Limited research has explored the robustness of a recommendation model in terms of fairness, e.g., the parity in performance across groups, under attack scenarios.
In this paper, we aim to assess the robustness of graph-based recommender systems concerning fairness, when exposed to attacks based on edge-level perturbations.
To this end, we considered four different fairness operationalizations, including both consumer and provider perspectives.
Experiments on three datasets shed light on the impact of perturbations on the targeted fairness notion, uncovering key shortcomings in existing evaluation protocols for robustness.
As an example, we observed perturbations affect consumer fairness on a higher extent than provider fairness, with alarming unfairness for the former.
Source code: \url{https://github.com/jackmedda/CPFairRobust}.
\end{abstract}

\section{Introduction} \label{sec:intro}
Individuals are increasingly interacting with recommender systems, enjoying the benefits of personalized services provided by e-commerce and streaming platforms.
These services are designed to adapt to the preferences and interests of consumers about the content they discover, while also meeting the expectations of content providers, who seek visibility and engagement.
However, the experiences of these stakeholders can be compromised by specialized attacks targeting recommender systems.
These attacks aim to manipulate the recommendations generated by the systems according to the attacker's objectives~\cite{DBLP:journals/corr/abs-2309-02057,DBLP:conf/aaai/OMahonyHS05,DBLP:reference/sp/AnelliDNM22}.

The effectiveness of attacks against recommender systems has been demonstrated across a diverse range of recommendation models, including those based on $k$-nearest neighborhood~\cite{DBLP:conf/aaai/OMahonyHS05}, matrix factorization~\cite{DBLP:conf/nips/LiWSV16}, association rules~\cite{DBLP:conf/ndss/YangGC17}, recurrent neural networks~\cite{DBLP:conf/icde/SongLHWLLG20}, and graphs~\cite{DBLP:conf/acsac/FangYGL18,DBLP:journals/tois/NguyenQNHVNJN23}.
In this context, `poisoning' attacks have become particularly prevalent~\cite{DBLP:journals/corr/abs-2309-02057,DBLP:journals/tkde/SunDYZWYHL23}.
These attacks primarily involve data perturbation during the training stage~\cite{DBLP:reference/sp/AnelliDNM22,DBLP:journals/tkde/SunDYZWYHL23}, often through the introduction of fake users, also known as 'shilling' attacks~\cite{DBLP:reference/sp/AnelliDNM22,DBLP:conf/aaai/OMahonyHS05,DBLP:conf/icde/SongLHWLLG20}.
Concerted efforts have been increasingly devoted to enhancing the robustness of recommender systems against various attacks~\cite{DBLP:journals/corr/abs-2309-02057,DBLP:conf/sigir/ChenZLHWY22,DBLP:conf/sigir/0001HDC18,DBLP:conf/sigir/ZhangYCHHC20,DBLP:conf/www/ZhangY0HC021}.
Fraudster detection~\cite{DBLP:conf/sigir/ZhangYCHHC20,DBLP:conf/sigir/CaoCY0Z20} and adversarial learning~\cite{DBLP:conf/sigir/ChenZLHWY22,DBLP:conf/sigir/0001HDC18,DBLP:conf/aaai/WuWWH021} have emerged as the primary defensive strategies.
The former seeks to identify and mitigate the influence of fake users, while the latter introduces perturbations to strengthen models against adversarial samples.

Differently from other research fields (e.g., computer sensing~\cite{DBLP:conf/icml/Croce020a,DBLP:conf/icml/CroceGBS0C22,DBLP:conf/nips/YinLSCG19,DBLP:conf/nips/BarbuMALWGTK19}, code generation~\cite{DBLP:conf/icse/MastropaoloPGCSOB23}, and program understanding~\cite{DBLP:conf/issta/ZengTZLZZ22}), attacks against recommendation have prioritized the maximization of model disruption, often at the expense of constructive objectives, without considering their impact on model robustness.
The existing methods for evaluating robustness merely compare recommendation utility before/after attacks~\cite{DBLP:journals/corr/abs-2309-02057}.
This practice is unfortunately limited, given that the overall utility can remain stable even when recommendations are significantly altered~\cite{DBLP:conf/cikm/OhUMK22}.
This limitation prevents from detecting the impact of attacks on beyond-accuracy objectives, such as trustworthiness~\cite{WangZWLR22}, fairness~\cite{WanngMZLM22}, and explainability~\cite{ZhangC20}.
Few prior works consider robustness in beyond-accuracy properties, such as bias~\cite{DBLP:conf/recsys/SatoTSO20} and sparseness~\cite{DBLP:conf/kdd/ZhengMGZ21}, but do not cover fairness~\cite{DBLP:journals/corr/abs-2309-02057}.
In other domains, this interplay has already been addressed~\cite{DBLP:conf/acl/PruksachatkunKD21,DBLP:journals/tifs/MarrasKJM23}.

In this paper, we provide a novel comprehensive analysis on the robustness of graph-based recommender systems in terms of fairness, referred to as \emph{robustness in fairness}.
Specifically, we investigate the extent to which the system fairness remains stable, from both the consumer~\cite{10.1007/978-3-030-99736-6_37,DBLP:journals/tois/0007D0F0023,WanngMZLM22} and the provider~\cite{DBLP:conf/kdd/SinghJ18,DBLP:conf/sigir/GomezZBSM21,DBLP:journals/fgcs/GomezZBSR22} sides, under attack scenarios.
We address this issue on systems based on graphs due to their state-of-the-art performance and the extensive range of attacks on graph data~\cite{DBLP:journals/tkde/SunDYZWYHL23,DBLP:conf/acsac/FangYGL18,DBLP:conf/www/ZhangY0HC021,DBLP:journals/tois/NguyenQNHVNJN23,DBLP:conf/sigir/ZhangYCHHC20,DBLP:conf/recsys/Christakopoulou19}.
{\color{black} Adding and deleting edges is a popular technique for attacks in graph data~\cite{DBLP:journals/tkde/SunDYZWYHL23}.}
To this end, we extended an approach that perturbs a graph at the edge-level to explain the predictions in several downstream tasks~\cite{DBLP:conf/aistats/LucicHTRS22,DBLP:conf/pkdd/KangLB21}.
This approach iteratively performs poisoning-like attacks against recommender systems based on Graph Neural Networks (GNNs) and monitors fairness as the user-item interaction graph gets gradually perturbed, encompassing different types of perturbations and fairness operationalizations.
{\color{black} Although our experimental evaluation is driven by the employed attack,  it is important to note that the attack itself does not constitute the main contribution.
Rather, this paper specifically aims to study the robustness in fairness in recommendation and explore the nuances in the GNN models' outcomes after~attacks.}

Our study operates within a white-box scenario, simulating the role of an attacker aiming to compromise the group fairness of a recommender system.
In this scenario, the perturbation process involves modifying the input graph that feeds into the GNN.
Such attacks may have real-world consequences, including compromising a company's reputation in the public eye~\cite{DBLP:conf/aaai/MehrabiNMG21}, both through media coverage and legal implications that may result in sanctions and other repercussions.
Concerning the recent regulations in terms of robustness and fairness of automated systems~\cite{DBLP:journals/cacm/NoiaTFS22,capAI}, such consequences could illustrate a worrying scenario.
Our experimental study showcases an extensive characterization of robustness in fairness against poisoning-like attacks, by employing three datasets and three GNN-based recommender systems.
The tested models exhibit a higher sensitivity to attacks tailored for group consumer fairness compared with provider one.
Specifically, the unfairness levels across consumer groups can be increased by a restrained amount of perturbations, whereas the impact on provider fairness is limited by the prior unfairness level exhibited in the original recommendations.

\section{Related Work} \label{sec:rel_works}

\subsection{Attacks and Robustness in Recommendation} \label{subsec:robust_recommendation}

The researchers addressing attacks and robustness in recommendation do not necessarily see these properties as interconnected, although most of the literature in robustness regards attacks~\cite{DBLP:journals/corr/abs-2309-02057}.
In fact, several papers solely focused on identifying attacks and treating them as strategies for achieving maximum disruptions~\cite{DBLP:conf/aaai/OMahonyHS05,DBLP:conf/acsac/FangYGL18,DBLP:conf/recsys/Christakopoulou19,DBLP:conf/icde/SongLHWLLG20,DBLP:conf/ndss/YangGC17,DBLP:conf/nips/LiWSV16} by injecting fake users to increase the recommendation of specific items~\cite{DBLP:conf/aaai/OMahonyHS05,DBLP:conf/acsac/FangYGL18}, or adversarially generate unnoticeable fake profiles~\cite{DBLP:conf/recsys/Christakopoulou19}.
Conversely, other works focused on improving robustness without necessarily viewing the adopted attack as the actual contribution~\cite{DBLP:conf/cikm/OhUMK22,DBLP:conf/sigir/ZhangYCHHC20,DBLP:conf/cikm/YuanYB20,DBLP:conf/sigir/ChenZLHWY22,DBLP:conf/recsys/YueZKSW22}.
For instance, \cite{DBLP:conf/sigir/ZhangYCHHC20} detected suspicious users as fraudsters using neural random forests.
{\color{black} Despite these advancements, a comprehensive analysis of the impact of attacks on the models' robustness, particularly in terms of accuracy and other critical properties, is notably absent.
This gap in research is remarkable, especially when compared to analogous studies conducted in other fields~\cite{DBLP:conf/icml/Croce020a,DBLP:conf/icml/CroceGBS0C22,DBLP:conf/icse/MastropaoloPGCSOB23,DBLP:conf/nips/YinLSCG19,DBLP:conf/nips/BarbuMALWGTK19,DBLP:conf/issta/ZengTZLZZ22,DBLP:conf/acl/PruksachatkunKD21}.}

\subsection{Fairness in Recommendation} \label{subsec:fairness_recommendation}

Due to recently issued regulations~\cite{DBLP:journals/cacm/NoiaTFS22,capAI}, researchers are increasingly prioritizing beyond-accuracy aspects in recommendation, as explainability~\cite{ZhangC20} and fairness~\cite{WanngMZLM22}.
The relevant amount of recent works studying consumer and provider (un)fairness addressed their assessment~\cite{10.1007/978-3-030-99736-6_37,DBLP:journals/ipm/BorattoFMM23,DBLP:conf/sigir/GomezZBSM21,DBLP:journals/fgcs/GomezZBSR22,DBLP:conf/kdd/SinghJ18}, mitigation~\cite{DBLP:conf/www/LiCFGZ21,DBLP:conf/fat/BurkeSO18,DBLP:conf/cikm/BorattoFFMM23,DBLP:conf/wsdm/GeLGXLZP0GOZ21,DBLP:journals/umuai/BorattoFM21}, and explanation~\cite{DBLP:conf/sigir/GeTZXL0FGLZ22,DBLP:conf/wsdm/GhazimatinBRW20,DBLP:journals/ipm/DeldjooBN21,DBLP:journals/corr/abs-2304-06182}.
{\color{black} Despite calls for unifying the goals of robustness and fairness in recommendation~\cite{DBLP:journals/corr/abs-2309-02057}, to the best of our knowledge, \cite{DBLP:conf/recsys/YangLZZC23} is the only work that focused on both properties. Specifically, \cite{DBLP:conf/recsys/YangLZZC23} proposed a fair and distributionally robust method to solve the distribution shift problem between the training and testing sets.}
Unfortunately, no study addressed the assessment of robustness in group consumer/provider fairness against specialized attacks.

\subsection{Robustness and Beyond-accuracy Aspects} \label{susbec:robust_beyond_acc}

Some studies on robustness in recommendation considered beyond-accuracy properties, e.g., bias~\cite{DBLP:conf/recsys/SatoTSO20} and sparseness~\cite{DBLP:conf/kdd/ZhengMGZ21}, as pertaining to a kind of robustness~\cite{DBLP:journals/corr/abs-2309-02057}.
However, their scope does not cover the fairness property envisioned in our study.
{\color{black} On the other hand, the literature in other fields has witnessed the introduction of novel techniques of certified robustness for text classification~\cite{DBLP:conf/acl/PruksachatkunKD21}, and novel attacks that target the fairness of classifiers~\cite{DBLP:conf/aaai/MehrabiNMG21,DBLP:conf/pkdd/SolansB020}.}
Nevertheless, their works regard classification tasks, where attacks and robustness methods differ from the ones employed in recommendation, as those targeted by this paper.

\section{Methodology} \label{sec:method}

\subsection{Perturbation Task in Graph-based Recommendation} \label{subsec:prob_form}

Our perturbation task tailored for GNN-based recommender systems aims to perturb the adjacency matrix through edge perturbations to alter the predicted recommendation lists, and test the systems' robustness in fairness.
We then distinguish between the recommendation task and the proper perturbation task.

\subsubsection{Recommendation task.} \label{subsubsec:rec_task}

In a typical recommendation scenario, a model learns the preferences of a set of users $U$ from their past interactions with a catalog of items $I$.
The network of user-item interactions can be represented by means of an undirected bipartite graph $G = (V, E)$, where $V = U \cup I$ is the set of nodes, $E$ is the set of edges between user and item nodes.
$G$ can be encoded in a $n \times n$ adjacency matrix $A$, where $A_{u,i} > 0$ denotes an edge links the user $u$ with the item $i$, otherwise $A_{u, i} = 0$.
We can feed $A$ to a GNN $f$ to predict the probability of missing user-item links.
Specifically, $f$ can be parameterized by a weight matrix $W$ and optimized to recommend to each user a list of the top-$k$ items sorted by the predicted linking probability in descending order.
Let $q_u@k$ be the top-$k$ list recommended to user $u$ and $Q@k$ the set of all $q_u@k, \forall u \in U$.

\subsubsection{Perturbation task.} \label{subsubsec:perturb_task}

Following~\cite{DBLP:journals/corr/abs-2309-02057}, robustness can be estimated by the disparity between the performance measured with the original (non-perturbed) data and the perturbed data.
A model reporting a disparity lower than a threshold $\epsilon$ and a perturbation bounded by a constant $\gamma$ would be denoted as ($\gamma$, $\epsilon$)-robust.
In our graph-based scenario, the original data is the adjacency matrix $A$, and we denote its perturbed version as $\tilde{A}$.
Given our fairness-related task, we define the performance by means of a fairness metric $M$.
We can then formally define the ($\gamma$, $\epsilon$)-robustness in fairness of our recommender system $f$ as follows:
\begin{equation} \label{eq:fair_robustness}
    \Delta = M(f(\tilde{A}, W), A) - M(f(A, W), A), \quad \left\Vert \Delta \right\Vert^2_2 \leq \epsilon, 
    \quad |\tilde{E}| \leq \gamma
\end{equation}
where $M$ estimates the fairness level based on the outcome of $f$ and $A$, and $\tilde{E}$ denotes the set of candidate edges for perturbation.
{\color{black} Although \cite{DBLP:journals/corr/abs-2309-02057} does not guide the selection of $\epsilon$ and $\gamma$, a small $\epsilon$ guarantees a greater level of robustness, while a small $\gamma$ reflects an attack that is harder to detect.}
Focused on analyzing the robustness in fairness, we do not set a fixed bound on the extent of edge perturbations, i.e. $\gamma = +\infty$.
Addressing edge perturbations as deletion and addition, the range of $|\tilde{E}|$ is then $[1, |E|]$ for deletion, and $[1, |U| \times |I| - |E|]$ for addition.

We seek to test the robustness in fairness in recommendation by identifying the edge perturbations that maximize the disparity in \eqref{eq:fair_robustness}, prioritizing fewer perturbations.
In other words, we aim to optimize the following objective function:
\begin{equation} \label{eq:long_objective_function}
    \min_{\hat{p}} \quad -\left\Vert M(f(\tilde{A}, W; \hat{p}), A) - M(f(A, W), A) \right\Vert^2_2 + \lambda\left\Vert \Gamma(\tilde{A}, A) \right\Vert^2_2
\end{equation}
where $\Gamma$ is a distance function~\cite{DBLP:journals/tkde/SunDYZWYHL23,DBLP:conf/aistats/LucicHTRS22},
$\hat{p}$ is a trainable weight used to identify the edges to be perturbed, $\lambda \in \mathbb{R}$ is a hyper-parameter that is used to control the weight between the two terms.
The minus sign applied to the first term optimizes the objective function towards maximizing the disparity $\Delta$, resulting in unfair recommendations.
However, the original fairness level estimated by $M$ is not affected by the perturbation process.
Hence, we can simplify \eqref{eq:long_objective_function} as follows:
\begin{equation} \label{eq:short_objective_function}
    \min_{\hat{p}} \quad -M(f(\tilde{A}, W; \hat{p}), A) + \lambda\left\Vert \Gamma(\tilde{A}, A) \right\Vert^2_2
\end{equation}

\subsection{Graph Perturbation Mechanism} \label{subsec:graph_perturb}

Following works of explainability in GNNs~\cite{DBLP:conf/aistats/LucicHTRS22,DBLP:conf/pkdd/KangLB21}, we use a sparsification method to obtain a binary perturbation tensor from a trainable real-valued weight (first described by \cite{DBLP:conf/cvpr/SrinivasSB17}) and extend it to the recommendation scenario.
First, we enlarge the space of the candidate edges to the entire graph.
Second, we replace the perturbation matrix $P$ used in \cite{DBLP:conf/aistats/LucicHTRS22} with a binary perturbation vector $p \in \{0, 1\}^{|\tilde{E}|}$, such that solely the relevant user-item connections are perturbed instead of affecting also self-loops, user-user and item-item links.
$p$ can be derived from the trainable weight $\hat{p}$ \eqref{eq:long_objective_function} by applying a sigmoid transformation and a binarization to $\hat{p}$, so as to map values lower than $0.5$ to $0$, otherwise to $1$.

The perturbation mechanism can be thought as a substitution process that updates the entries of the adjacency matrix $A$ with the entries of $p$, resulting in $\tilde{A}$.
A fixed relation between the 2D index of $A$ and the 1D index of $p$ establishes which candidate edge $(u, i) \in \tilde{E}$ will be perturbed by the $j$-th entry of $p$.
The entries update process resulting in $\tilde{A}$ is formally defined as:
\begin{equation} \label{eq:perurbation}
    \tilde{A} = A \dotplus p, \quad
    \begin{aligned}
        \tilde{A}_{u, i} = 
        \begin{cases}
            p_j & \text{if} \; (u, i) \in \tilde{E}  \\
            A_{u, i} & \text{otherwise}
        \end{cases}
    \end{aligned}
\end{equation}
$\dotplus$ denotes the perturbation operator for edge deletion \emph{$\dotplus$ Del} or addition \emph{$\dotplus$ Add}.

This perturbation mechanism is performed iteratively by gradually modifying $A$ to generate $\tilde{A}$, until the perturbed edges optimize the targeted task.
Specifically, we initialize $\hat{p}$ based on the perturbation type (deletion or addition), such that no edge is affected in $A$, i.e. $A \dotplus p = A$.
At each iteration, we generate $\tilde{A}$ with $\hat{p}$, and feed $\tilde{A}$ to the GNN $f$ to produce the corresponding recommendations.
The latter are processed by \eqref{eq:short_objective_function} to estimate the fairness level, the distance between $\tilde{A}$ and $A$, and to update the weight $\hat{p}$ accordingly.
Finally, the process stops based on a predefined criterion, e.g., the impact of the last perturbation.

\subsection{Fairness Notion and Operationalization} \label{subsec:fair_notion_and_operat}

\subsubsection{Fairness Notion.} \label{subsubsec:fair_notion}

We proceed to define the fairness metric $M$.
We follow recent works~\cite{10.1007/978-3-030-99736-6_37,10.1145/3564285,DBLP:conf/www/LiCFGZ21,DBLP:conf/sigir/GeTZXL0FGLZ22,DBLP:conf/kdd/SinghJ18,DBLP:conf/wsdm/GeLGXLZP0GOZ21,DBLP:conf/interspeech/FenuMMM21,DBLP:conf/icb/AtzoriFM22} that emphasized the relevance of the group fairness notion of \emph{demographic parity} from both the consumer~\cite{10.1007/978-3-030-99736-6_37,10.1145/3564285,DBLP:conf/www/LiCFGZ21} and provider side~\cite{DBLP:conf/sigir/GeTZXL0FGLZ22,DBLP:conf/kdd/SinghJ18,DBLP:conf/wsdm/GeLGXLZP0GOZ21}.
For the former side, demographic parity is satisfied if consumer groups experience the same level of recommendation utility.
For the latter side, the notion is satisfied if the probability of being recommended is equal across provider groups, proportionally to their representation in the catalog.

We ground our work on a binary setting as previous studies~\cite{10.1007/978-3-030-99736-6_37,DBLP:conf/www/LiCFGZ21,DBLP:conf/sigir/GeTZXL0FGLZ22,DBLP:conf/wsdm/GeLGXLZP0GOZ21}, where each stakeholder set $Z$ ($Z \subseteq U$ for consumers, $Z \subseteq I$ for providers\footnote{We do not consider other features and associate each item with a distinct provider.}) can be partitioned in two groups, $Z_1$ and $Z_2$.
Multiple attributes, e.g., gender and age, of each stakeholder could produce a distinct partition of $Z$ in two groups.
Demographic parity (DP) can then be operationalized as the following disparity:
\begin{equation} \label{eq:dp}
    DP = \left\Vert S(f(A, W), A^{Z_1}) - S(f(A, W), A^{Z_2}) \right\Vert^2_2
\end{equation}
where $S$ represents the metric used to estimate the performance w.r.t. the corresponding stakeholder, e.g., exposure for a provider, $A^{Z_1}$ and $A^{Z_2}$ respectively denote the adjacency sub-matrices with regard to the two partitions $Z_1$ and $Z_2$.

\subsubsection{Consumer and Provider Fairness Operationalization.} \label{subsubsec:consumer_provider_operat}

Depending on the adopted metric $S$, we can define specific operationalizations of DP, which reflect distinct perspectives of the unfairness issue.
For each stakeholder, we contemplate two types of operationalization, a rank-aware and a rank-agnostic one.

We underline that DP represents $M$ in \eqref{eq:fair_robustness}-\eqref{eq:long_objective_function}-\eqref{eq:short_objective_function}, given that it estimates the fairness performance of a recommender system.
Therefore, we define each operationalization of DP, the corresponding metric $S$ used for evaluation, and the differentiable approximation of $S$ to be used in the objective function in \eqref{eq:short_objective_function}:
\begin{itemize}
    \item \textbf{Consumer Preference} (CP): it estimates the consumer fairness as the disparity across consumers groups in rank-aware top-$k$ recommendation utility, which can be measured by the Normalized Discounted Cumulative Gain (NDCG@$k$).
    Following \cite{10.1145/3564285,DBLP:journals/ir/QinLL10}, NDCG (N@$k$) is approximated as the differentiable function $\widehat{NDCG}$, where the rank of an item is defined in terms of the pairwise preference with respect to any other item in the catalog.
    \item \textbf{Consumer Satisfaction} (CS): it estimates the consumer fairness as the disparity across consumers groups in rank-agnostic top-$k$ recommendation utility, which can be measured by the Precision (P@$k$).
    We optimize P@$k$ by treating the recommendation task as a binary classification task, by using a sigmoid function followed by a binary cross entropy loss.
    This task aims to include relevant items in the top-$k$ list, regardless of their position.
    \item \textbf{Provider Exposure} (PE): it estimates the provider fairness as the disparity in exposure across providers groups.
    Following \cite{DBLP:conf/sigir/GomezZBSM21,DBLP:journals/fgcs/GomezZBSR22,DBLP:conf/sigir/GeTZXL0FGLZ22,DBLP:conf/wsdm/GeLGXLZP0GOZ21}, we define the exposure of a generic provider group $I_*$ as the average number of exposures in $Q@k$ (estimated by an indicator function $\mathbb{1}[\cdot]$ as \cite{DBLP:conf/sigir/GeTZXL0FGLZ22,DBLP:conf/wsdm/GeLGXLZP0GOZ21}), discounted by the importance of their position~\cite{DBLP:conf/kdd/SinghJ18,DBLP:conf/sigir/GomezZBSM21,DBLP:journals/fgcs/GomezZBSR22} (as for DCG), and normalized by the ideal exposure~\cite{DBLP:conf/sigir/GomezZBSM21,DBLP:journals/fgcs/GomezZBSR22} (as for NDCG).
    Formally:
    \begin{equation} \label{eq:pe}
        \text{Exposure}(I_* \; | \; Q@k) = \frac{|I|}{|I_*|} \; \frac{1}{|U|} \sum_{u \in U}{\frac{\sum^k_{j = 1}{\frac{\mathbb{1}[i_j \in I_*]}{log_2(j + 1)}}}{\sum^k_{j = 1}{\frac{\mathbb{1}[i_j \in I]}{log_2(j + 1)}}}}
    \end{equation}
    We leverage the approximation proposed by \cite{DBLP:conf/sigir/GeTZXL0FGLZ22}, where the indicator function is replaced by the predicted linking probability (item relevance).
    \item \textbf{Provider Visibility} (PV): it estimates the provider fairness as the disparity in visibility across providers groups.
    Following \cite{DBLP:conf/icwsm/FabbriCB022,DBLP:conf/sigir/GomezZBSM21,DBLP:journals/fgcs/GomezZBSR22}, we define the visibility of a generic provider group $I_*$ as the average number of exposures in $Q@k$ (estimated by $\mathbb{1}[\cdot]$ as for PE).
    Formally:
    \begin{equation} \label{eq:pv}
        \text{Visibility}(I_* \; | \; Q@k) = \frac{|I|}{|I_*|} \; \frac{1}{|U|k} \sum_{u \in U}{\sum_{i \in q_u@k}{\mathbb{1}[i \in I_*]}}
    \end{equation}
    We approximate PV by replacing $\text{Visibility}(\cdot, \cdot)$ with the loss function used for $CS$, which would aim to include the items of $I_*$ in the top-$k$ list.
\end{itemize}

CP and CS $\in [0, 1]$, PE and PV $\in [0, \frac{|I|}{|I_1|}]$, for which $0$ denotes fairness.

The approximations of PE and PV are inspired by \cite{DBLP:conf/sigir/GeTZXL0FGLZ22}, but the authors measure the disparity only according to the top-$k$ items, limiting the exposure/visibility information, given that $k$ is usually small.
It causes the gradient to be computed only for the top-$k$ items, which will not necessarily be included in the final recommendations, due, for instance, to items already enjoyed by some users.
Instead, we set $k$ to be $10\%$ of the items catalog size $|I|$ to expand the operational scope of PE and PV.
This choice also helps when just one of the groups is represented in $Q@k$, and enables our perturbation task to better optimize the presence of one of the groups in positions closer to the top-$k$ ones.

\section{Experimental Evaluation}

The following experiments aim to answer the following research questions:
\begin{enumerate}[start=1,label={\bfseries RQ\arabic*:},leftmargin=1.1cm]
    \item What is the extent to which edge perturbations impact the robustness in fairness of recommender systems?
    \item Are the adopted models similarly affected in terms of alterations in robustness in fairness as edge perturbations gradually increase?
    \item Which consumer or provider group should be more affected by the edge perturbation to nuke the models’ robustness in fairness?
\end{enumerate}

\subsection{Evaluation Setting} \label{subsec:eval_settings}

\subsubsection{Evaluation Protocol}

The perturbation process is run for 200 epochs, but early stopped if the increment in $\Delta$ after 15 consecutive epochs is lower than $0.001$.

Given our objective of testing the robustness in fairness, we do not perform a classic poisoning attack as defined in \cite{DBLP:journals/corr/abs-2309-02057,DBLP:reference/sp/AnelliDNM22,DBLP:journals/tkde/SunDYZWYHL23}.
Specifically, the combination of datasets, models, perturbation types, and fairness operationalizations sums up to 108 attacks, but we aim to also address intermediary perturbation stages, which would result in an impracticable amount of models re-training processes.
To this end, we estimate the first term of $\Delta$ by substituting the original adjacency matrix $A$ with the perturbed one $\tilde{A}$ at the inference stage, and maintaining the models' parameters constant.
If $\tilde{A}$ was generated by external tools or approaches, such poisoning-like attack would reflect a white/grey-box setting, with the unique requirement of having access to the saved representation of $A$.

\subsubsection{Models.}

We rely on Recbole~\cite{DBLP:conf/cikm/ZhaoMHLCPLLWTMF21} and select GCMC~\cite{DBLP:journals/corr/BergKW17}, LightGCN (LGCN)~\cite{DBLP:conf/sigir/0001DWLZ020}, and NGCF~\cite{DBLP:conf/sigir/Wang0WFC19} as the GNN-based recommender systems for our study.
{\color{black} Though the set of employed models is limited, they cover different architectures to learn the users' preferences: GCMC leverages an auto-encoder structure, LGCN learns from linear relationships between users and items, NGCF adopts features transformation and nonlinear activation on the message-passing step.}
While the last two models generate the recommendations by the learnt user and item embeddings, GCMC performs a complete forward process during inference.
Thus, GCMC represents a more suitable candidate for the attacker, who does not need to force the embeddings re-generation after the graph perturbation.

\begin{table}[!t]
    \renewcommand{\arraystretch}{1.2}
    \caption{\textbf{Left}: datasets' statistics (\emph{Repr}: \emph{Representation}, \emph{O}: \emph{Older}, \emph{Y}: \emph{Younger}, \emph{F}: \emph{Female}, \emph{M}: \emph{Males}). \textbf{Right}: original models' performance in N@$10$ (\%) and P@$10$ (\%).}
    \label{tab:datasets_models}
    \vspace{-2mm}
    \begin{minipage}{.5\linewidth}
      \centering
        \resizebox{1\linewidth}{!}{
            \begin{tabular}{rr r|r|r}
            \toprule
                   &    &  ML1M~\cite{DBLP:journals/tiis/HarperK16} &  LF1K~\cite{DBLP:books/daglib/0025137} & INS~\cite{DBLP:conf/www/LiCFGZ21} \\
            \cline{3-5}
            \multicolumn{2}{r}{\# Users \;}&                    6,040 & 268      & 346 \\
            \multicolumn{2}{r}{\# Items \;}&                    3,706 & 51,609   & 20 \\
            \multicolumn{2}{r}{\# Interactions \;}&         1,000,209 & 200,586  & 1,879 \\
            \multicolumn{2}{r}{Domain \;}&                      Movie &    Music & Insurance \\
            \cline{3-5}
            \multirow{4}{*}{Repr.} & \multirow{2}{*}{Age} &   O : 43.4\%  &   O : 42.2\%  & O : 49.4\% \\
            & & Y : 56.6\% & Y : 57.8\% & Y : 50.6\% \\
            \cline{2-5}
            & \multirow{2}{*}{Gender} & F : 28.3\%  &   F : 42.2\% & F : 23.4\% \\ 
            & & M : 71.7\% & M : 57.8\% & M : 76.6\% \\
            
            \bottomrule
            \end{tabular}
            }
    \end{minipage}%
    \begin{minipage}{.5\linewidth}
      \centering
        \resizebox{0.9\linewidth}{!}{
        \begin{tabular}{l|rr|rr|rr}
            \toprule
            & \multicolumn{2}{c}{INS} & \multicolumn{2}{c}{LF1K} & \multicolumn{2}{c}{ML1M} \\
            & N@10 & P@10 & N@10 & P@10 & N@10 & P@10 \\
            \midrule
            GCMC & 76.40 & 9.60 & 39.66 & 38.47 & 12.62 & 11.35 \\
            \midrule
            LGCN & 78.07 & 9.74 & 39.81 & 38.36 & 12.68 & 11.35 \\
            \midrule
            NGCF & 78.39 & 9.74 & 39.71 & 38.43 & 12.94 & 11.65 \\
            \bottomrule
        \end{tabular}
        }
    \end{minipage}
\vspace{-2mm}
\end{table}

\subsubsection{Datasets.}

We rely on \cite{10.1007/978-3-030-99736-6_37}, which includes MovieLens-1M~\cite{DBLP:journals/tiis/HarperK16} (ML1M) and LFM-1K~\cite{DBLP:books/daglib/0025137} (LF1K)\footnote{The timestamp of each $(u,i)$ refers to the last interaction between $u$ and a $i$'s song.}.
We also consider Insurance~\cite{DBLP:conf/sigir/LiCXGZ21} (INS) and discard consumers with less than 5 interactions.
Table~\ref{tab:datasets_models} reports the datasets' statistics and the origial models' performance.
We use the items' popularity to partition the provider set in short-head $I_1$ and long-tail $I_2$ items such that $\frac{|I_1|}{|I_2|} = \frac{1}{4}$ as in \cite{DBLP:conf/sigir/GeTZXL0FGLZ22,DBLP:conf/wsdm/GeLGXLZP0GOZ21}.
For each user, the interactions are sorted in ascending order of recency, split by a ratio 7:1:2, and each split respectively assigned to the training, validation, and testing set.
We use the validation set to select the best original model, and the ground truth of the testing set to optimize the fairness operationalizations (Section~\ref{subsec:fair_notion_and_operat}).
Using the test relevance judgements would result in a powerful attack and support our main concern, i.e. the analysis of robustness in fairness\footnote{This design choice does not constitute the issue highlighted in \cite{DBLP:conf/sigir/RahmaniNDA22}.}.

\subsection{RQ1: Impact on Robustness in Fairness} \label{subsec:rq1}

\begin{figure}[!t]
    \centering
    \begin{tabular}{p{0.2cm}c}
         \multicolumn{2}{c}{\includegraphics[width=0.75\linewidth]{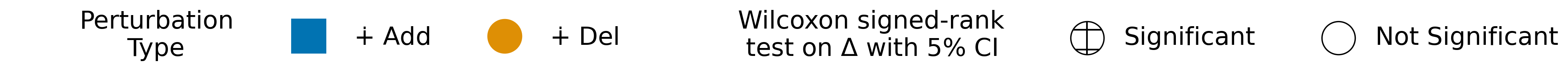}} \\
         \multirow{1}{*}[9mm]{\rotatebox[origin=c]{90}{\scriptsize INS}} & \includegraphics[width=0.97\linewidth]{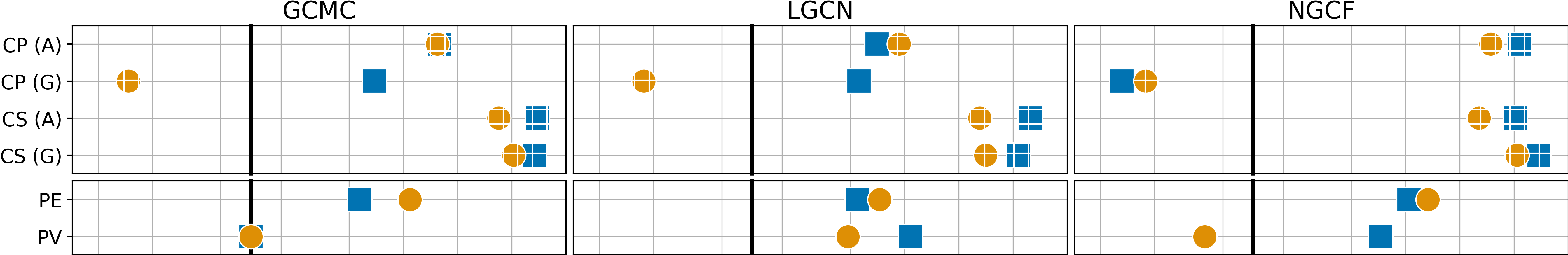} \\
         \multirow{1}{*}[12mm]{\rotatebox[origin=c]{90}{\scriptsize LF1K}} & \includegraphics[width=0.97\linewidth]{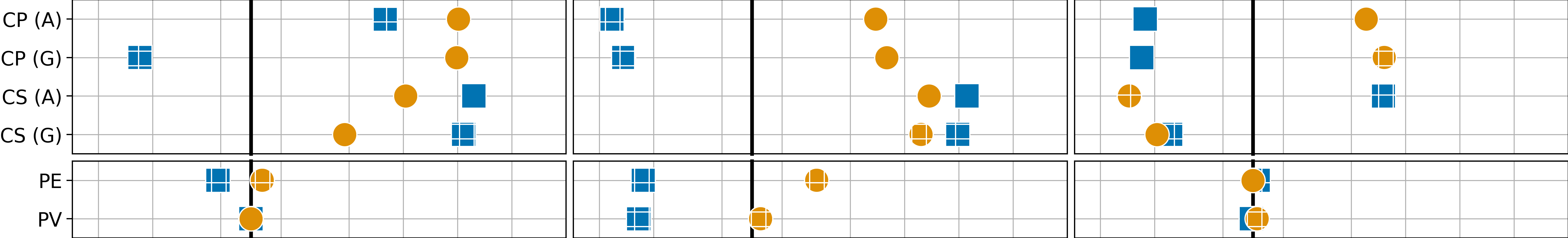} \\
         \multirow{1}{*}[13mm]{\rotatebox[origin=c]{90}{\scriptsize ML1M}} & \includegraphics[width=0.97\linewidth]{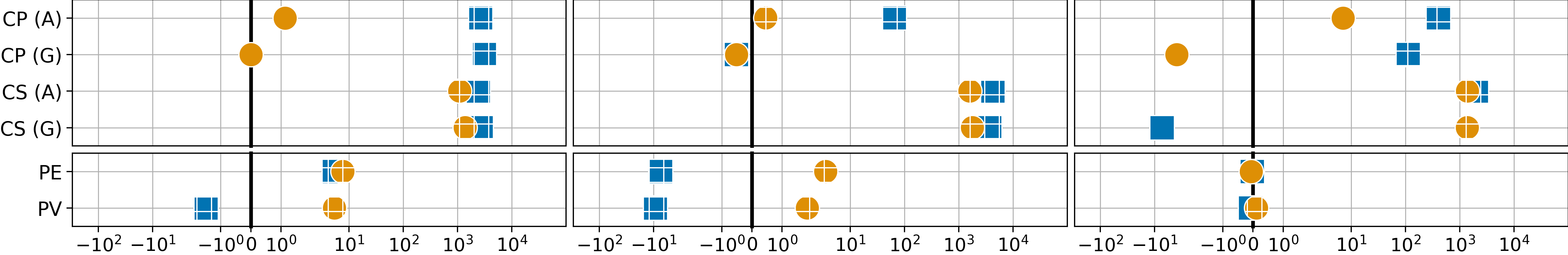} \\
    \end{tabular}
    \caption[Impact of Edges Perturbation on Robustness in Fairness]{Impact of edges addition (\emph{$\dotplus$ Add}) and deletion (\emph{$\dotplus$ Del}) on the robustness in fairness, reported as the relative difference in $M$ between the non-perturbed and perturbed model, i.e. $\Delta/M(f(A, W), A)$ (\eqref{eq:fair_robustness}). \emph{A} stands for \emph{Age}, \emph{G} for \emph{Gender}.}
    \label{fig:robustness_impact}
    \vspace{-2mm} 
\end{figure}

We first assess the impact of edge-level perturbations on the robustness in fairness, i.e. $\Delta$, where $M$ corresponds to the formulation of DP in \eqref{eq:dp} and its specialized operationalizations, e.g., PE for provider exposure fairness.

Figure~\ref{fig:robustness_impact} reports the impact on robustness in fairness as the relative difference in $M$ between the original fairness level and after each perturbation type, i.e. $\Delta$ divided by the second term in \eqref{eq:fair_robustness}.
The reported values pertain to the iteration where at least one edge was perturbed and $\tilde{A}$ affected the most $\Delta$ for each model.

The formulation of $\Delta$ in \eqref{eq:fair_robustness} enables us to easily denote positive values as increment in fairness level disparity across groups, i.e. increment in DP, otherwise a decrement of the latter.
Therefore, positive values reflect a successful outcome for the attacker, whose goal is to make the system generate unfair recommendations.
{\color{black} Negative values derive from instances of $\tilde{A}$ that reduce the unfairness level even at the first iterations.}
The x-axis has been symmetrically log-normalized to highlight the remarkable impact on $\Delta$ on the systems.
For instance, both \emph{$\dotplus$ Add} and \emph{$\dotplus$ Del} significantly caused an impact for CS on LGCN under ML1M, precisely increasing DP by more than 1,000\%.
Also the fairness levels under INS were remarkably affected, reaching values higher than 10,000\%.

Successful attacks are especially reported in terms of edge deletions, given that most of the points labeled as \emph{$\dotplus$ Del} are depicted at the right of the zero line.
Conversely, \emph{$\dotplus$ Add} caused varied results, ranging from reductions of fairness level disparity, e.g., on LGCN under LF1K, to disruptions in robustness in fairness, e.g., on GCMC under ML1M for the consumer side.
{\color{black} This observation could be related to the addition of information in the graph due to \emph{$\dotplus$ Add}, but also to the} larger edge sample space for the latter compared with \emph{$\dotplus$ Del}.
The experiments on provider fairness confirm \emph{$\dotplus$ Del} as a more effective perturbation attack than \emph{$\dotplus$ Add}.
However, several settings report an optimal robustness, e.g., GCMC and NGCF under LF1K, and other ones a negative orientation of $\Delta$, e.g., LGCN under LF1K and ML1M.
{\color{black} Across the models, NGCF exhibits the least sensitivity to perturbations, especially on provider fairness and under LF1K, given that most of the points are close to the zero line.
This may be attributed to the feature transformation and nonlinear activation applied in the NGCF message-passing scheme, which diminish the impact of the perturbed graph on the predictions.}

\subsection{RQ2: Robustness in Fairness under Incremental Perturbations} \label{subsec:rq2}

\begin{figure}[!t]
    \centering
    \begin{tabular}{p{0.3cm}l}
         & \includegraphics[width=0.9\linewidth]{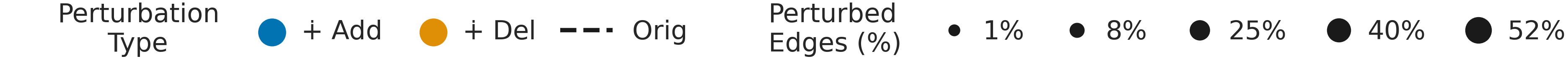} \\[4pt]
         \multirow{1}{*}[14.5mm]{\rotatebox[origin=c]{90}{INS}} & \includegraphics[width=0.97\linewidth]{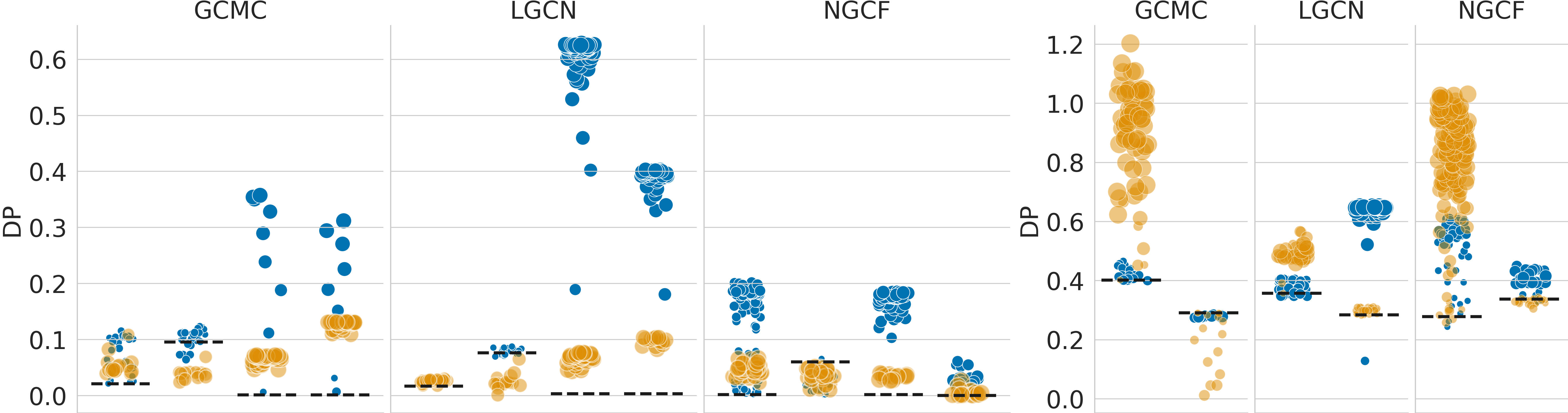} \\[2pt]
         \multirow{1}{*}[17mm]{\rotatebox[origin=c]{90}{LF1K}} & \includegraphics[width=0.97\linewidth]{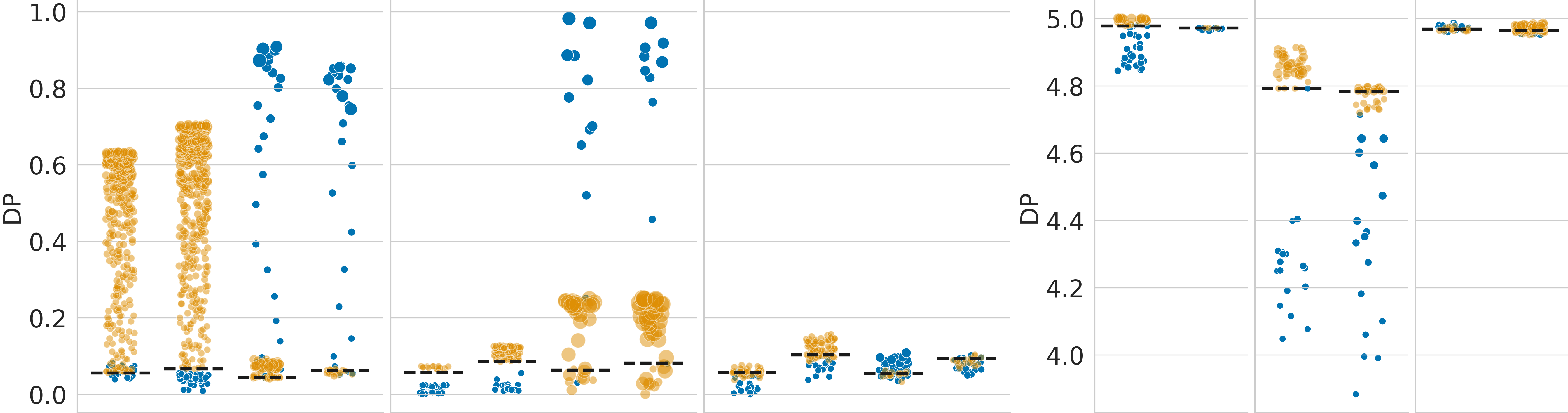} \\[2pt]
         \multirow{1}{*}[20mm]{\rotatebox[origin=c]{90}{ML1M}} & \includegraphics[width=0.97\linewidth]{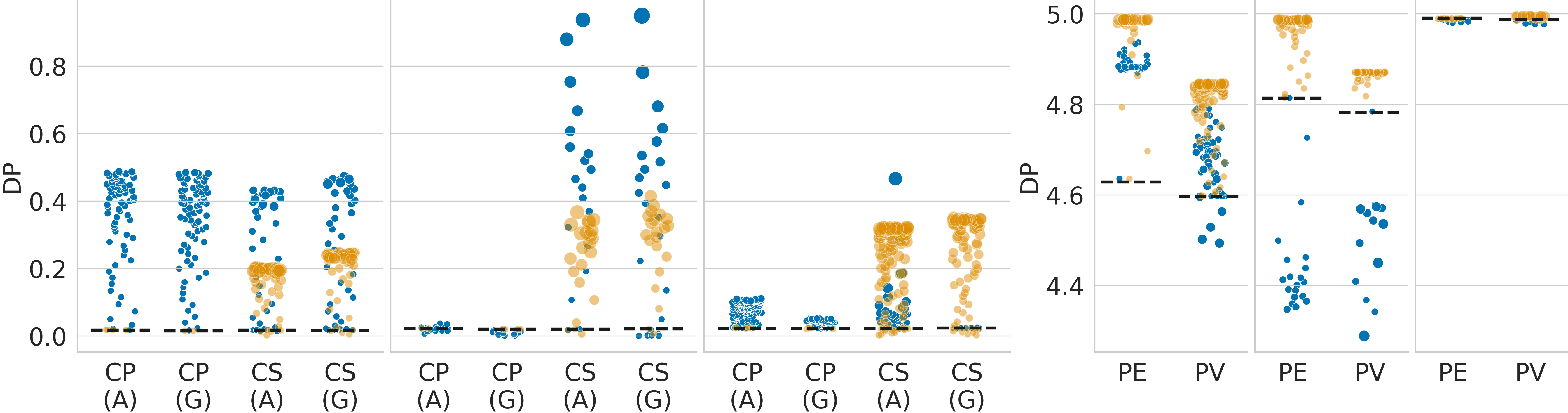} \\
    \end{tabular}
    \vspace{-3mm} 
    \caption{Trend of the operationalizations of DP across different stages of the perturbation process. Each stage reflects a fraction of perturbed edges w.r.t. $\tilde{E}$, depicted as points gradually larger as the amount of perturbed edges increases. The horizontal dashed line labeled as \emph{Orig} denotes the DP of the recommendations generated by the non-perturbed system. In the consumer-side results, \emph{A} stands for \emph{Age}, \emph{G} for \emph{Gender}.}
    \label{fig:dp_over_pert_edges}
\end{figure}

The previous research question aimed to highlight if the edge perturbations could be able to affect the robustness in fairness of the considered models, accounting only for the iteration with the highest impact on $\Delta$.
It is then unclear to what extent the robustness is affected by the gradual increment of edge perturbations.

To this end, we report the DP for each iteration of the perturbation process through the points in Figure~\ref{fig:dp_over_pert_edges}.
For each setting, a horizontal dashed line denotes the original DP to highlight the gap caused by the increment of perturbed edges.
A robust model would then be described by points close to the dashed line, while far points conceive the perturbed edges significantly affected the fairness level.
Even the size of such points is important: big points close to the dashed line denote the model is robust despite a relevant amount of perturbed edges.

\vspace{-3mm} \paragraph{Consumer-side.}

Across the datasets, NGCF is the most robust model, given that in most of the setting the points are distributed closer to the original DP compared with the other two systems.
{\color{black} This likely stems from the observation highlighted in the previous section.}
GCMC and LGCN behave in a similar way, especially in terms of CS, for which the addition of edges (\emph{$\dotplus$ Add}) has a more significant influence on DP compared with the deletion (\emph{$\dotplus$ Del}).
However, LGCN reports a systematic robustness for the rank-aware operationalization CP with only a few small points reported in the proximity of the original DP.

Some experiments underline the attack could be influenced by the dataset and sensitive attribute.
The former influence factor can be observed on GCMC, for which \emph{$\dotplus$ Del} significantly affected CP under LF1K; a similar result was reported under ML1M, but due to the other perturbation type \emph{$\dotplus$ Add}.
The latter influence factor can be observed under INS on NGCF, where the impact of \emph{$\dotplus$ Add} on DP is reported only across age groups, regardless of the operationalization.

\vspace{-3mm} \paragraph{Provider-side.}
Differently from the consumer-side, the original systems report a high degree of unfairness, close to the maximum value $\frac{|I|}{|I_1|} = 5$.
Hence, the edge perturbations adopted by a potential attacker cannot significantly affect the robustness in fairness, as highlighted under LF1K on all models, and under ML1M on NGCF, where the original level of PE and PV is remarkable.

In the other settings, deleting edges is more reliable in influencing the robustness in provider fairness compared with \emph{$\dotplus$ Add}.
This behavior is especially reported under INS, but also under ML1M on GCMC and LGCN. 

Other observations contrast with the consumer-side evaluation.
First, NGCF exhibits a high degree of sensitivity to the perturbations under INS, as emphasized by the gradual increment in PE as more edges are deleted.
{\color{black} This is possibly due to the limited size of INS, which prevents NGCF from robustly learning the users' preferences in a generalized way.}
Second, the differences in trend between GCMC and LGCN are more remarked, with \emph{$\dotplus$ Add} affecting more the robustness of GCMC under ML1M w.r.t. the LGCN's one.
{\color{black} This can be explained by the GCMC encoder architecture, which possibly interprets the added edges as noise, which is instead captured as new information by the linear step in LGCN.}

\subsection{RQ3: Edge Perturbations Influence on Groups} \label{subsec:rq3}

\begin{figure}[!b]
    \centering
    \begin{tabular}{p{0.3cm}l}
         & \includegraphics[width=0.97\linewidth]{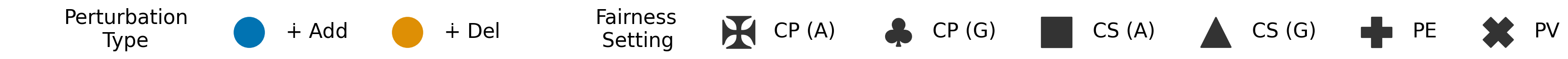} \\
         & \scriptsize \hspace{3cm} Consumer-side \hspace{4cm} Provider-side \\[3pt]
         \multirow{1}{*}[30.5mm]{\rotatebox[origin=c]{90}{\scriptsize $\Delta EI$}} & \includegraphics[width=0.97\linewidth]{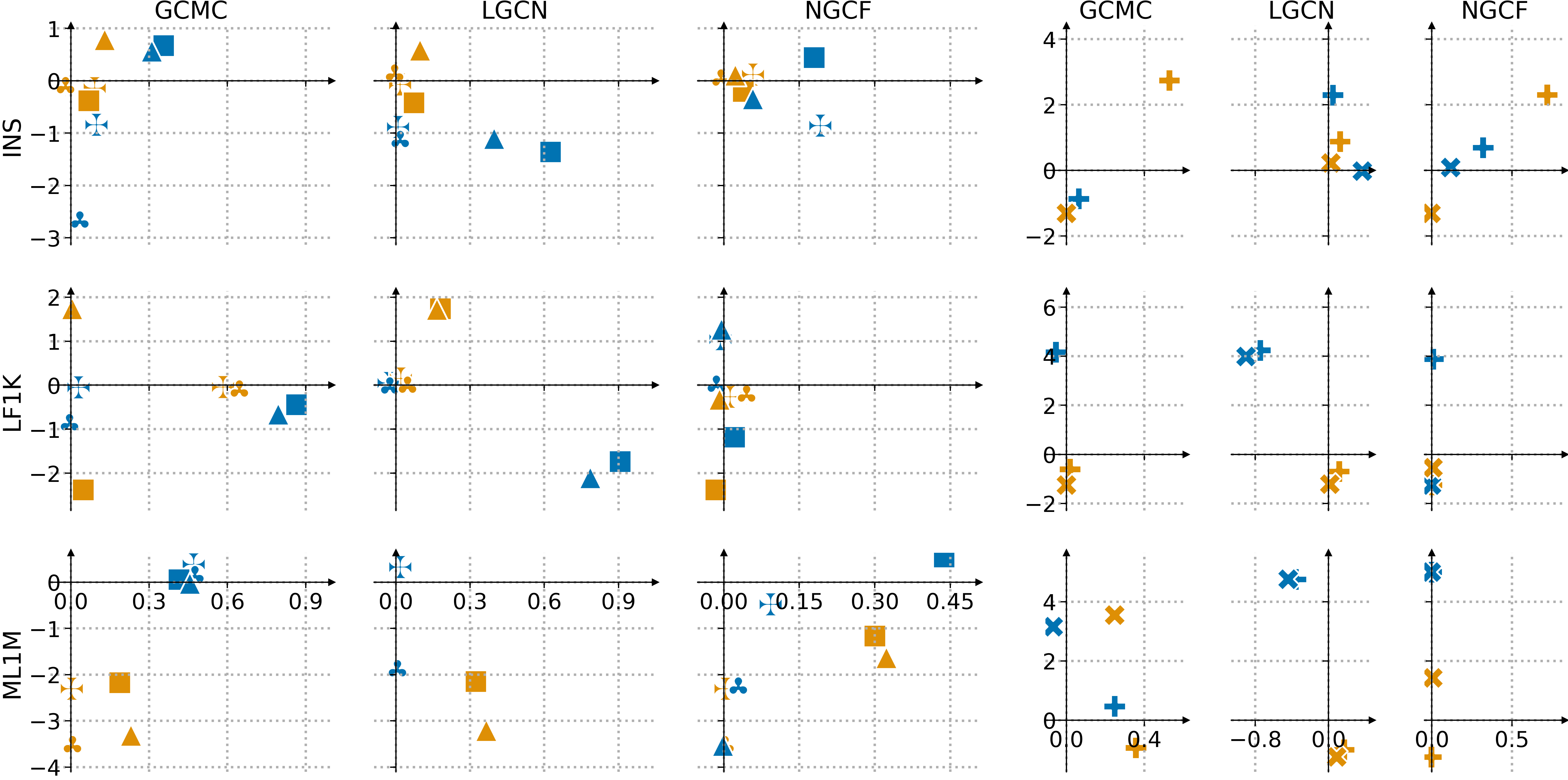} \\[0pt]
         & \hspace{5.3cm} $\Delta$
    \end{tabular}
    \caption{Relationship between $\Delta EI$ (y-axis), i.e. the disparity in edge perturbations distribution between the advantaged and the disadvantaged group, and $\Delta$ (x-axis), i.e. the disparity in fairness level before and after the attack. Successful attacks ($\Delta > 0$) are caused by targeting more the advantaged group if $\Delta EI > 0$, otherwise the disadvantaged group was targeted. \emph{A} stands for \emph{Age}, \emph{G} for \emph{Gender}. Some points overlap.}
    \label{fig:deltaei_vs_delta}
    \vspace{-5mm}
\end{figure}

Group unfairness denotes one of the groups is favored by the system, e.g., with a higher level of exposure.
We denote such group as \emph{advantaged} and the other group as \emph{disadvantaged}.
We seek to discover whether the impact on $\Delta$ is due to a greater amount of edge perturbations applied to the advantaged or the disadvantaged group.
To this end, we define \emph{Edge Impact} (EI) as the ratio between the distribution of edge perturbations on the nodes of a group and the representation of the latter.
Let $\neg\tilde{E} \subseteq \tilde{E}$ be the subset of edges actually perturbed, formally $EI_* = \frac{|\neg\tilde{E}_*|/|\neg\tilde{E}|}{|Z_*|/|Z|}$,
where $Z_*$ is a group of consumers or providers, and $\neg\tilde{E}_*$ is the subset of edges perturbed w.r.t. the nodes in $Z_*$.
We also define the difference $\Delta EI = EI_{Adv} - EI_{Disadv}$ between the advantaged and the disadvantaged group.

In Figure~\ref{fig:deltaei_vs_delta}, we depict $\Delta EI$ on the y-axis and $\Delta$ on the x-axis to study their relationship.
The values of $\Delta$ pertain to our first experiment in Section~\ref{subsec:rq1}.

In the consumer side, most of the settings reporting an impact on $\Delta$ were caused by edge perturbations prioritizing the disadvantaged group, i.e. $\Delta EI < 0$.
In particular, this behavior regards \emph{$\dotplus$ Del}, which removed interactions of the disadvantaged group to, intuitively, reduce its recommendation utility.
Conversely, we expect \emph{$\dotplus$ Add} to report $\Delta EI > 0$.
However, this is actually true only under ML1M on LGCN and NGCF, and slightly under INS on GCMC and NGCF.
Other settings, e.g., under LF1K, illustrate $\Delta$ was affected by adding more edges to the disadvantaged group, highlighting that a higher number of interactions does not systematically correspond to a higher recommendation utility.

In the provider side, the disparity in edge perturbations distribution is more remarked in comparison with the consumer-side results.
However, such disparity does not reflect a relevant impact on $\Delta$ in most settings, as already highlighted in other experiments.
Successful attacks ($\Delta > 0$) are mostly reported under INS for PE, where $\Delta EI$ is systematically above the origin.
Hence, the impact on the robustness in provider fairness was caused by deleting or adding more edges to the advantaged group.
{\color{black} Other scenarios where $\Delta$ was affected, e.g., on GCMC under ML-1M, confirm such observation or report close points to the x-axis.}

\section{Conclusions} \label{sec:conclusions}

In this paper, we present the concept of robustness in fairness and raised attention towards the issues caused by related attacks in recommendation. Compared to prior work, our analysis aimed to assess the robustness in fairness of GNN-based recommender systems against poisoning-like attacks based on edge-level perturbations, focusing on the models' robustness and not on the attack itself.

{\color{black} Even though the range of considered models is limited, they represent consistent baselines in the literature, and cover varied GNN architectures. It follows that the adopted models represent perfect candidates for an analysis of robustness in fairness, a topic that is still unexplored in recommendation compared with other fields.} Nevertheless, future works will cover a wider set of models. 

Moreover, although the adopted attack is focused to test the robustness in fairness, the same approach could be extended to targeted attacks, e.g., against specific consumers or providers. In other words, based on how each stakeholder partitions are established, our attack could be used to modify the model performance to specifically favor one of the tailored groups. We plan to investigate other types of attack, such as those relying on perturbations based on re-wiring.

Additionally, our white-box setting could be simplified to a grey-box one by removing the assumption of having access to the model parameters. Given that the attack solely requires the substitution of the adjacency matrix with its perturbed version, the latter could be generated through a different process. Future work will be focused on exploring grey- or black-box settings.

\vspace{2mm} \noindent \textbf{Acknowledgement}.
We acknowledge financial support under the National Recovery and Resilience Plan (NRRP), Miss. 4 Comp. 2 Inv. 1.5 - Call for tender No.3277 published on Dec 30, 2021 by the Italian Ministry of University and Research (MUR) funded by the European Union – NextGenerationEU. Prj. Code ECS0000038 eINS Ecosystem of Innovation for Next Generation Sardinia, CUP F53C22000430001, Grant Assignment Decree N. 1056, Jun 23, 2022 by the MUR.

\bibliographystyle{splncs04}
\bibliography{sample}

\end{document}